\documentclass[%
 reprint,
 superscriptaddress,
 amsmath,amssymb,
 aps,
prl
floatfix,
]{revtex4-2}
\usepackage{graphicx}
\usepackage{dcolumn}
\usepackage{bm}
\usepackage[utf8]{inputenc}
\usepackage[T1]{fontenc}
\usepackage{etoolbox}
\usepackage{calc}
\usepackage{siunitx}
\usepackage{comment}
\usepackage{graphicx}
\usepackage[shortlabels]{enumitem}
\usepackage{amsmath}
\usepackage{amsfonts}
\usepackage{amssymb}
\usepackage{float}
\usepackage{xcolor}
\usepackage[version=4]{mhchem}
\usepackage{spalign}
\usepackage{systeme}
\usepackage{physics}
\usepackage{epsfig}
\usepackage{color}

\begin{document}

\title{Testing the Renormalization of the von Klitzing Constant by Cavity Vacuum Fields}
\author{Josefine Enkner}
\affiliation{Institute of Quantum Electronics, ETH Zürich, 8093 Zürich, Switzerland}
\affiliation{Quantum Center, ETH Zürich, 8093 Zürich, Switzerland}
\author{Lorenzo Graziotto}
\affiliation{Institute of Quantum Electronics, ETH Zürich, 8093 Zürich, Switzerland}
\affiliation{Quantum Center, ETH Zürich, 8093 Zürich, Switzerland}
\author{Felice Appugliese}
\affiliation{Institute of Quantum Electronics, ETH Zürich, 8093 Zürich, Switzerland}
\affiliation{Quantum Center, ETH Zürich, 8093 Zürich, Switzerland}
\author{Vasil~Rokaj}
\affiliation{ITAMP, Center for Astrophysics $|$ Harvard $\&$ Smithsonian, MA 02138 Cambridge, USA}
\affiliation{Department of Physics, Harvard University, MA 02138 Cambridge, USA}
\author{Jie~Wang}
\affiliation{Department of Physics, Harvard University, MA 02138 Cambridge, USA}
\affiliation{Center of Mathematical Sciences and Applications, Harvard University, MA 02138 Cambridge, USA\looseness=-1}
\affiliation{Department of Physics, Temple University, Philadelphia, Pennsylvania, 19122, USA}
\author{Michael~Ruggenthaler}
\affiliation{Max Planck Institute for the Structure and Dynamics of Matter, 22761 Hamburg, Germany}
\author{Christian Reichl}
\affiliation{Laboratory for Solid State Physics, ETH Zürich, 8093 Zürich, Switzerland}
\author{Werner Wegscheider}
\affiliation{Laboratory for Solid State Physics, ETH Zürich, 8093 Zürich, Switzerland}
\author{Angel~Rubio}
 \affiliation{Max Planck Institute for the Structure and Dynamics of Matter, 22761 Hamburg, Germany}
\affiliation{Center for Computational Quantum Physics, Flatiron Institute, NY 10010 New York, USA}
\author{Jér\^{o}me Faist}
\affiliation{Institute of Quantum Electronics, ETH Zürich, 8093 Zürich, Switzerland}
\affiliation{Quantum Center, ETH Zürich, 8093 Zürich, Switzerland}

\begin{abstract}
In light of recent developments demonstrating the impact of cavity vacuum fields inducing the breakdown of topological protection in the integer quantum Hall effect, a compelling question arises: what effects might cavity vacuum fields have on fundamental constants in solid-state systems? In this work we present an experiment that assesses the possibility of the von Klitzing constant itself being modified. By employing a Wheatstone bridge, we precisely measure the difference between the quantized Hall resistance of a cavity-embedded Hall bar and the resistance standard, achieving an accuracy down to 1 part in $10^{5}$ for the lowest Landau level. While our results do not suggest any deviation that could imply a modified Hall resistance, our work represents pioneering efforts in exploring the fundamental implications of vacuum fields in solid-state systems.
\end{abstract}
\maketitle

In condensed matter physics the vacuum fields refer to the quantum fluctuations of the ground-state electromagnetic field, which owing to Heisenberg's uncertainty principle possess zero average but finite variance~\cite{milonni1994quantum}. Although energy conservation prevents the direct probing of their virtual excitations, vacuum fields manifest their physical reality via the interaction with matter systems, leading to experimentally accessible modifications of their energy spectra, such as the Lamb shift~\cite{lambShift}, or fundamental properties, such as the electron magnetic moment~\cite{electrongfactor}. Both effects are very well understood within quantum electrodynamics (QED)~\cite{cohen1989photons, peskin1995introduction}. In recent years the possibility of shaping the vacuum fields inside cavities has been proposed~\cite{cavitronics, hubener2021engineering} as a means of engineering matter properties, such as the electron-phonon coupling~\cite{EPCcavity} and the molecular structure~\cite{cavityMoleculeStrong}, altering matter phases, such as superconductivity~\cite{schlawin2019cavity} and ferroelectricity~\cite{cavityFerroelectric}, or affecting non-equilibrium phenomena like chemical reactions~\cite{ribeiro2018polariton} and charge transport~\cite{cavityTransport}. Experimental progresses have been made particularly in showing the cavity-altered ground-state chemical reactivity~\cite{thomas2016ground, ahn2023modification}, and recently we experimentally showed how cavity vacuum fields can destroy the topological protection of the integer quantum Hall states~\cite{appugliese2022breakdown}. The breakdown manifests in the finite longitudinal resistivity of the otherwise zero-resistance edge states, accompanied by a loss of quantization of the transverse (Hall) resistivity. These effects are explained by the long-range perturbation introduced by cavity vacuum fields back-scattering the edge states~\cite{ciuti2021cavity}, in particular those at high odd integer filling factors $\nu = n_s h / e B$, where $n_s$ indicates the two-dimensional (2D) electron density, $h$ is Planck's constant, $e$ the electron charge, and $B$ the magnetic flux density in the direction orthogonal to the 2D electron system. Increasing the magnetic field, i.e.\ going to lower integer filling factors, quantization is fully recovered, with the longitudinal resistivity going to zero and the Hall resistivity developing plateaux. In 2D electron systems not embedded in cavity vacuum fields the value of these plateaux is given by reciprocal-integer multiples of von Klitzing's constant $R_K = h / e^2$, the equality being supported by theoretical gauge~\cite{laughlin1981quantized, PhysRevB.25.2185} and topological~\cite{PhysRevLett.49.405, thouless1994topological} arguments, and experimentally verified down to $8$ parts in $10^{11}$~\cite{wheatstoneBridge}. For this reason, the integer quantum Hall effect (IQHE)~\cite{klitzing1980new} remains essential in the foundation of the standard electrical resistance even after the changes in the International System of Units~\cite{proc-cgpm, rigosi2019quantum}. On the other hand, a precise measurement of the quantized values of the Hall plateaux for 2D electron systems embedded in cavities has not yet been performed, and one may inquire if these values show deviations from the value of $R_K$, that is if cavity vacuum fields can modify the value of $R_K$. Indeed, a tiny $10^{-20}$ correction due to the coupling to the free electromagnetic field was already calculated within QED~\cite{penin_hall_2009}, and it has been recently proposed~\cite{rokaj2022polaritonic} that the quantized Hall resistance of a cavity-embedded 2D electron system should depend on the light-matter coupling constant $\eta$ as $R_H = (1 + \eta^2) h/e^2 \nu$, where $\nu$ is the integer filling factor of the Hall plateau. In this work we show experimentally that cavity vacuum fields bear no impact on the value of the quantized Hall resistance down to $1$ part in $10^5$ for $\nu = 1$. \\



\begin{figure}
    \centering
    \includegraphics[width=\linewidth]{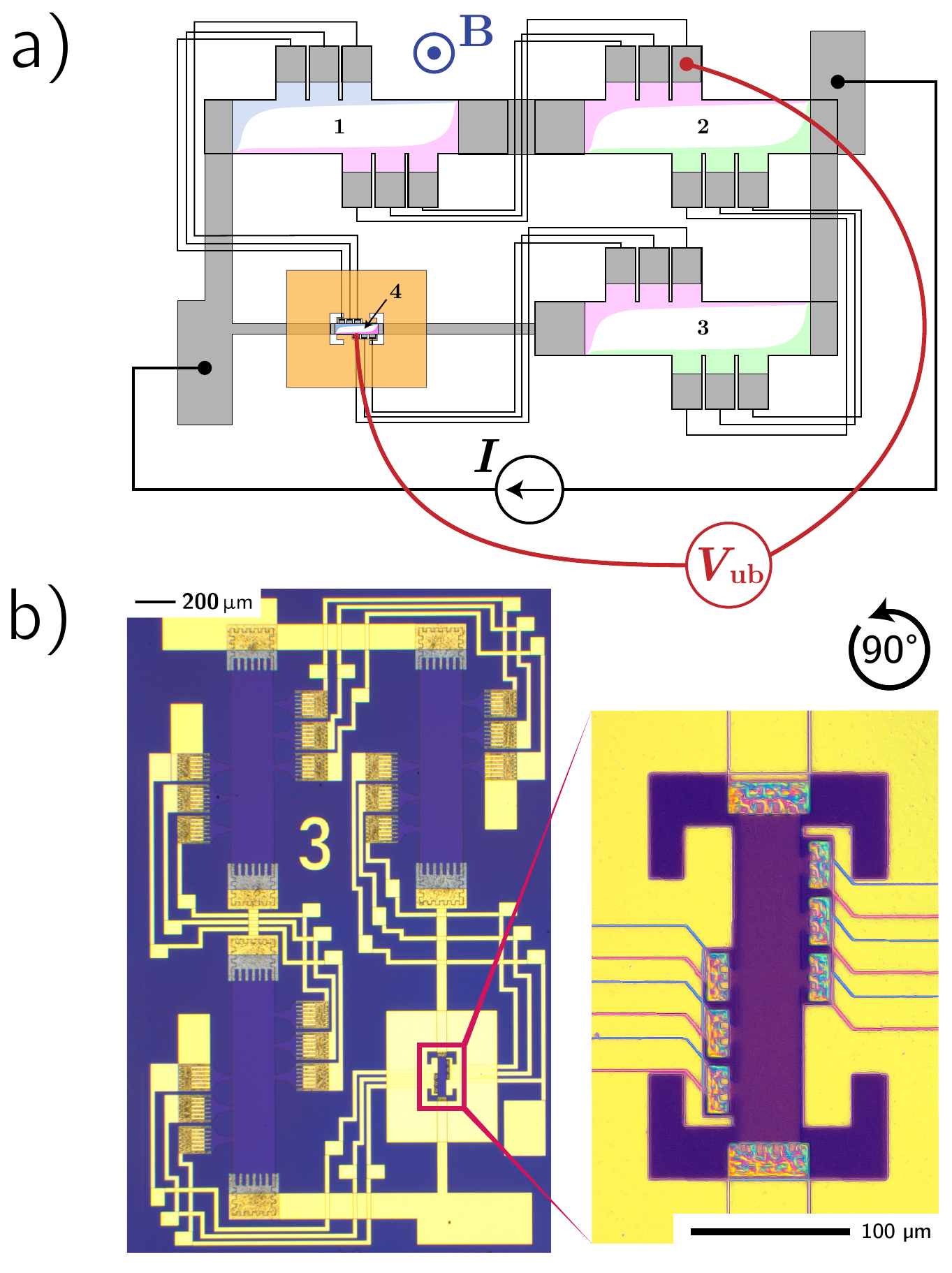}
    \caption{Layout of the sample. \textbf{a)} Sketch of cavity sample S3, where the measurement scheme is illustrated. The current $I$ is injected and extracted from the source-drain contacts, and the unbalance voltage $V_\mathrm{ub}$ is measured at the indicated voltage probes. The grey color indicates metal contacts and leads. The pale color shadings indicate regions having the same electric potential, in the situation in which the cavity does not alter the value of the quantized Hall resistance, so that the same current circulates in each arm of the Wheatstone bridge. The thin black lines indicate additional gold leads which connect contacts on regions of the Hall bars having the same electric potential, in order to ensure their equilibration. \textbf{b)} Microscope picture of cavity sample S3, with cavity-embedded Hall bar on the lower right, which is enlarged on the right. Notice that the picture is rotated by $90^\circ$ counter-clockwise with respect to a.}
    \label{fig:figure1_sample}
\end{figure}

The geometry of the experiment is centered around a $\SI{40}{\micro m}$ wide Hall bar placed in the capacitor gap of a complementary split-ring resonator (CSRR)~\cite{scalari2012ultrastrong} resonant at $\omega_\mathrm{cav}=2\pi \times \SI{140}{GHz}$, which we will refer to as the cavity in the following (see Figure~\ref{fig:figure1_sample}a, where the cavity-embedded Hall bar is labelled by number 4). CSRRs have the ability to strongly enhance vacuum fields $\mathcal{E} = \sqrt{\hbar \omega_{\rm cav}/2\epsilon_0 \epsilon_s V_\mathrm{eff}}$ in a sub-wavelength volume $V_\mathrm{eff}=\SI{2.7e5}{\micro m^3}$, $\epsilon_0$ being the vacuum permittivity, and $\epsilon_s=12.69$ the effective permittivity of GaAs, the material in which the 2D electron system is hosted (more details are provided in the Supplemental Material (SM)).  As a result, the interaction between enhanced cavity vacuum fields and the electrons occupying the last filled Landau level is pushed into the ultra-strong coupling regime~\cite{frisk2019ultrastrong}, leading to a normalized coupling $\Omega_R/\omega_{\rm cav}\approx0.3$, with $\Omega_R$ denoting the Rabi frequency~\cite{PhysRevB.81.235303}. We point out that in the experiment the Hall bar, together with the contacts to the 2D electron system, is completely inside the spatial gap of the cavity, as can be seen in Fig.~\ref{fig:figure1_sample}b, where a microscope picture of the sample and a zoom in on the cavity-embedded Hall bar are shown. One reason we do so is to suppress the scattering processes arising due to the vacuum field gradients at the boundary between the cavity capacitor gap and the etched lateral sides of the Hall bar. These field gradients amplify the effect that breaks the topological protection of the edge states. Here we are less interested in the vacuum fields role as a scattering source, but rather in whether the homogeneous component of vacuum fields may renormalize the value of fundamental constants. Therefore, we want to preserve the quantization of the Hall resistance, and assess whether the quantization value is changed. Another reasoning for the design follows from the following argument~\cite{laughlin1999nobel}:  all systems which are adiabatically connected (i.e.\ by varying some external parameters they can be transformed into one another without closing their spectral gap) to any system exhibiting a topological invariant must then also --- at zero temperature --- have that same topological invariant. The quantized transverse conductance $e^2\nu/h$ is a topological invariant for the IQHE. Since the predicted modification of the transverse conductance due to cavity vacuum fields is not necessarily topological in nature, putting the whole Hall bar structure, together with its contacts, inside the cavity prevents the system to relapse to its exactly quantized state at the contacts lying outside of the cavity. At the same time, two gapped systems with different Hall conductances must have a gap closing between them, a requirement which is fulfilled via the metal contacts connecting the Hall bars both in series and in parallel. Again, even if the deviation from the quantized value introduced by cavity vacuum fields in the Hall bar is not a topological invariant, we can preserve and access it experimentally. \\

To highlight deviations in standard Hall quantization compared to a Hall bar in the presence of cavity vacuum fields, we arrange three $\SI{200}{\micro m}$ wide bare Hall bars (i.e.\ not embedded in a cavity, see Fig.~\ref{fig:figure1_sample}a, where they are labelled by numbers 1-3) and the aforementioned $\SI{40}{\micro m}$ wide cavity-embedded Hall bar (number 4) in a Wheatstone bridge configuration~\cite{wheatstoneBridge}. 
This allows for the measurement of unknown resistances by detecting the unbalance voltage $V_\mathrm{ub}$ at the junction between the two resistors of each arm (see Fig.~\ref{fig:figure1_sample}a). By using three bare Hall bars as reference resistors, we take advantage of the fact that at the plateaux they possess the same resistance (i.e.\ a reciprocal-integer multiple of $R_K$), while the deviation of the cavity-embedded Hall bar resistance from the quantized value will be proportional to the unbalance voltage via
\begin{equation}\label{eq:res_dev}
    \frac{\Delta R_H}{R_H} = 4 \frac{\langle V_\mathrm{ub} \rangle}{I} \frac{e^2}{h} \nu,
\end{equation}
where $I = \SI{10}{nA}$ is the source-drain current that flows in the Wheatstone bridge, and $\langle V_\mathrm{ub} \rangle$ is the unbalance voltage averaged at the integer plateau $\nu$ (a derivation of Eq.~\ref{eq:res_dev} is given in the SM). The low injection current is in contrast to standard metrological measurement methods, since here we aim at achieving very low base temperatures in order to stay as close as possible to the regime for which the renormalization is predicted (close to $T=0$)~\cite{rokaj2022polaritonic}. At the plateaux, in the case in which no deviation due to cavity vacuum fields occurs, the voltage drop across both resistors in each arm will be the same, thus the unbalance voltage will be zero (see Fig.~\ref{fig:figure1_sample}a, where the pale color shadings indicate regions with equal electric potential in the case of no deviation). The finite contact resistance $R_C$ of the metallic contacts to the 2DES could hinder the measurement of $V_\mathrm{ub}$, hence the quadruple connection technique is employed~\cite{multipleConnections}. This technique involves introducing additional connections between pairs of Hall bars (shown by the three thin black lines in Fig.~\ref{fig:figure1_sample}a, in addition to the connection via the leads depicted in grey), each possessing a resistance $R_C$, which bridge the equi-potential sides of the Hall bars. We therefore ensure the equilibration of the equi-potential sides as the relative contribution of the contact resistance of each ohmic contact is reduced to a factor of $(R_C/R_H)^4$ (i.e.\ to about $1.4\times10^{-7}$ for filling factor $1$). To assess potential variations in the measurement result coming from sample-specific properties (due e.g.\ to inhomogeneities in the material), we fabricated one chip comprising 5 samples. Three of them (labelled S3, S4 and S5) possess the cavity-embedded Hall bar resonator, as discussed above (a microscope picture of Sample S3 is displayed in Fig.~\ref{fig:figure1_sample}b). Samples S1 and S2 have instead the same configuration but no resonator (only the bare \SI{40}{\micro m} Hall bar indicated by number 4 in Fig.~\ref{fig:figure1_sample}a) and serve as references. All 5 samples are processed from the same 2D electron system obtained in a GaAs/AlGaAs square quantum well, possessing an electron density of $n_s= \SI{2e11}{cm^{-2}}$ and a mobility of $\mu=\SI{1.6e7}{cm^2 V^{-1} s^{-1}}$. Indeed, the impact of cavity vacuum fields on transport has been shown on high-mobility samples~\cite{appugliese2022breakdown}, although such high values would be in contrast to standard metrological techniques. Fabrication is performed via standard photo-lithography, with an intermediate step where we have deposited an insulating layer of \ce{Al2O3} that electrically separates the overlapping gold planes and lines from each other. Vias in the insulating layer are opened with \ce{HF} etching, in order to achieve the quadruple connection scheme. \\

\begin{figure}
    \centering
    \includegraphics[width=1.\linewidth]{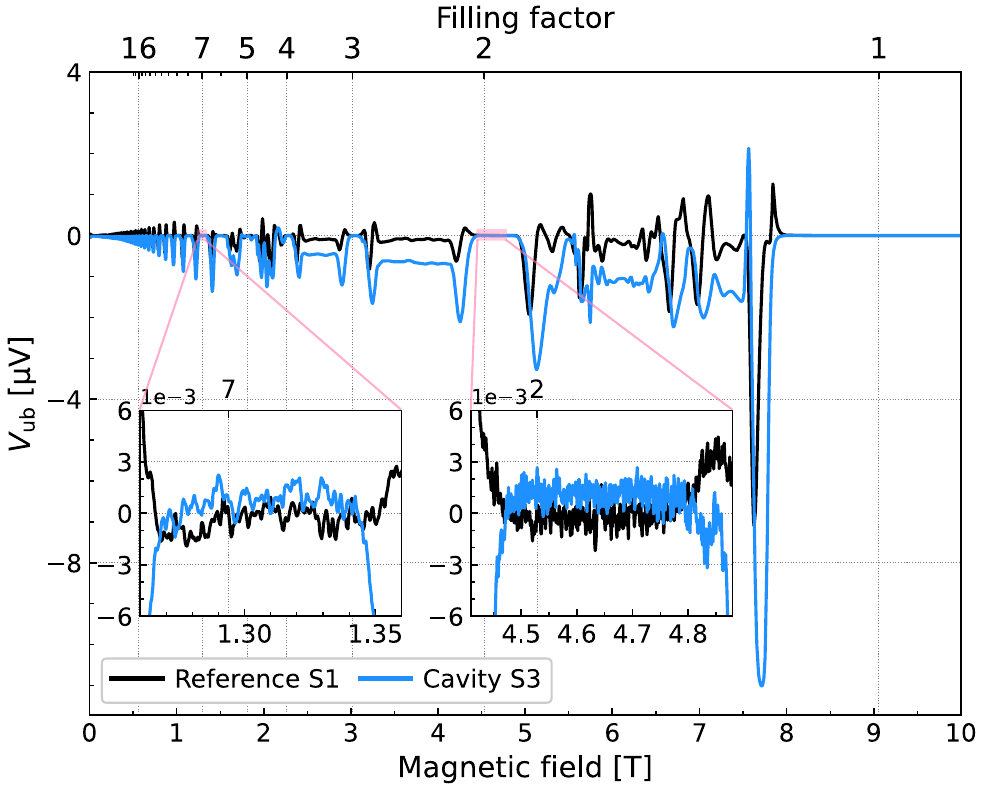}
    \caption{Measurement of the unbalance voltage of the Wheatstone bridge as a function of magnetic field for nominal value of the current injected $I=\SI{10}{nA}$, in the case of a sample without the cavity (black curve, S1) and with the cavity (blue curve, S3). The insets show a zoom near the plateaux $\nu=7$ and $\nu=2$. Notice that the vertical axis values of the insets are multiplied by $10^3$.}
    \label{fig:figure2_unbalV}
\end{figure}
\begin{figure}[hbt!]
    \centering
    \includegraphics[width=1.\linewidth]{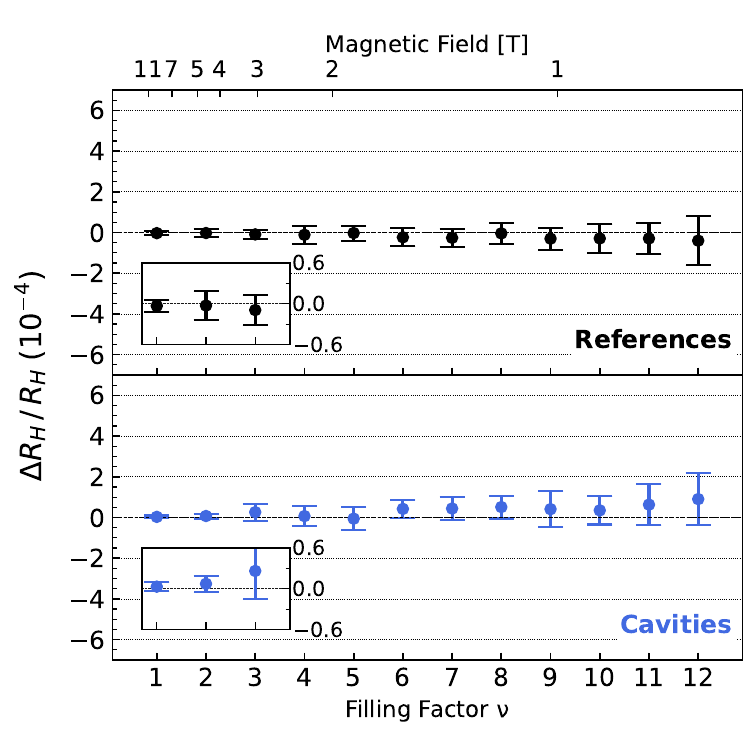}
    \caption{Weighted averages of the relative deviation from the Hall resistance as a function of the integer filling factor, for samples without the cavity (top panel) and with the cavity (bottom panel). The error-bars indicate a 99.7\% confidence interval of the values at the plateaux (i.e.\ three standard deviations). The relative deviation is below 1 part in $10^5$ for $\nu = 1$, as it can be seen in the insets, where the region between $\nu=1$ and $\nu=3$ is zoomed in, and it is below \SI{1e-4}{} for $\nu=12$.}
    \label{fig:figure3_plateauxAvg}
\end{figure}

In Figure~\ref{fig:figure2_unbalV} we display the unbalance voltage evolution as a function of perpendicular magnetic field $B$ for a reference (S1) and for a cavity (S3) sample at an electronic temperature below \SI{20}{mK}. We remind that a meaningful comparison between the two can only be performed at the plateaux, since away from them the three reference Hall bars have not a quantized resistance, and the unbalance voltage is mostly determined by the material inhomogeneities over the sample area. Zooming in on the plateaux regions (insets of Fig.~\ref{fig:figure2_unbalV}), we observe that both reference and cavity traces oscillate around zero, within a range of \SI{\pm 2}{nV}. The main contribution to the noise comes from the AC pre-amplifier voltage noise, which for a demodulation frequency of \SI{2.333}{Hz} amounts to about \SI{2.6}{nV / \sqrt{Hz}} (more details are given in the SM). The plateaux regions are identified by the filling factor $\nu$, and for the purposes of the following analysis their extension has been determined by taking \SI{3.2}{nV} as threshold value for the absolute value of $V_\mathrm{ub}$. 
In Figure~\ref{fig:figure3_plateauxAvg} we report the relative deviation $\Delta R_H / R_H$ from the Hall resistance at the integer plateaux as given by Eq.~\ref{eq:res_dev}, where we estimated the unbalance voltage as a weighted average for both reference and cavity samples. Indeed, the main source of uncertainty is statistical for $\nu > 2$, while for $\nu \leq 2$ sample-specific uncertainties dominate, and we assess them by measuring multiple reference and cavity samples (a more detailed discussion about the nature of the uncertainties is given in the SM). The error-bars display a 99.7\% confidence interval (i.e.\ their length is equal to three standard deviations) of the data distribution at each plateau. We notice that both the cavity and the reference samples show a deviation from the quantized Hall resistance compatible with zero within this confidence interval. 
Hence we reject the possibility of a relative deviation from the quantized value of the plateaux larger than three standard deviations, that is below $\SI{1e-5}{}$ for $\nu=1$, and below \SI{2e-4}{} for $\nu=12$. For higher filling factors (that is lower magnetic fields) the relative uncertainty is larger due to the smaller Hall resistance and the smaller extension of the plateaux. Although the precision we achieve is far from the one reported in Ref.~\cite{wheatstoneBridge}, it constitutes an upper bound to the effect that cavity vacuum fields may bear on the value of $R_K$. This result may be compared to the ratio of the Lamb shift to the Dirac levels of the hydrogen atom, which amounts to about \SI{4e-7}{}, and which quantifies the relative correction introduced by free vacuum fields to the energy levels uncoupled from the electromagnetic field, or to the relative contribution of about $10^{-3}$ of the radiative corrections on the magnetic moment of the electron~\cite{milonni1994quantum}, or also to the small $10^{-20}$ free-field radiative correction to $R_K$ predicted in Ref.~\cite{penin_hall_2009} within QED. Notice that in Ref.~\cite{rokaj2022polaritonic} a renormalized value was found in the limit of zero temperature and zero cavity frequency, while in our experiment the \SI{140}{GHz} cavity frequency cannot be assumed to be negligible. As shown in the SM, an approach based on the Kubo formalism~\cite{PhysRevB.98.205301, rokaj2023topological} shows indeed that for low enough temperatures (i.e.\ below the excitation energy of the light-matter hybrid states) but with finite cavity frequency one recovers the exact quantization of the Hall conductance in a homogeneous 2D electron system coupled to cavity vacuum fields. If the cavity frequency is considered to be negligible as compared to the other energy scales in the system, the lower polariton mode goes to zero (see also SM). This is a singular point for the system because the lower polariton gap closes, and thus circumvents the Thouless flux insertion argument which assumes a finite gap~\cite{thouless1994topological, rokaj2019}. The gap closing leads to the deviation of the Hall conductance from the precise quantization $\sigma_H=\nu e^2/h$~\cite{rokaj2022polaritonic, rokaj2019}. This singular point is not accessible in the regime of the current experiments.     \\

The present work serves as a foundation for investigating whether cavity vacuum fields cause any deviation in the Hall resistance from the von Klitzing constant. We have assessed that no deviation is present up to one part in $10^5$ for filling factor 1 by comparing a Hall bar embedded in a cavity to the standard resistance using the Wheatstone bridge measurement technique. In a more advanced version of this experiment, a further increase in precision can be achieved by injecting direct current, or by employing a cryogenic current comparator, hence investigating whether the deviation is smaller in magnitude, or if it arises due to finite temperature effects, as proposed by recent theoretical results~\cite{rokaj2023topological}. The possibility of affecting the value of fundamental constants via cavity vacuum fields could then have implications for many other solid-state systems. \\

We thank Giacomo Scalari for useful discussions, and Peter Märki for technical support. We acknowledge funding from the Swiss National Science Foundation (SNF) through grant no.\ 200020-207795. V.R.\ acknowledges support from the NSF through a grant for ITAMP at Harvard University. J.F.\ thanks the  Alexander von Humboldt stiftung (AvH) for their support.\\

J.E.\ and L.G.\ contributed equally to this work.

\bibliography{bibliography}

\end{document}


\onecolumngrid

\title{\textsc{Supplemental Material} \\[0.5cm] Testing the Renormalization of the von Klitzing Constant by Cavity Vacuum Fields}
\author{Josefine Enkner}
\affiliation{Institute of Quantum Electronics, ETH Zürich, 8093 Zürich, Switzerland}
\affiliation{Quantum Center, ETH Zürich, 8093 Zürich, Switzerland}
\author{Lorenzo Graziotto}
\affiliation{Institute of Quantum Electronics, ETH Zürich, 8093 Zürich, Switzerland}
\affiliation{Quantum Center, ETH Zürich, 8093 Zürich, Switzerland}
\author{Felice Appugliese}
\affiliation{Institute of Quantum Electronics, ETH Zürich, 8093 Zürich, Switzerland}
\affiliation{Quantum Center, ETH Zürich, 8093 Zürich, Switzerland}
\author{Vasil~Rokaj}
\affiliation{ITAMP, Center for Astrophysics $|$ Harvard $\&$ Smithsonian, MA 02138 Cambridge, USA}
\affiliation{Department of Physics, Harvard University, MA 02138 Cambridge, USA}
\author{Jie~Wang}
\affiliation{Department of Physics, Harvard University, MA 02138 Cambridge, USA}
\affiliation{Center of Mathematical Sciences and Applications, Harvard University, MA 02138 Cambridge, USA\looseness=-1}
\affiliation{Department of Physics, Temple University, Philadelphia, Pennsylvania, 19122, USA}
\author{Michael~Ruggenthaler}
\affiliation{Max Planck Institute for the Structure and Dynamics of Matter, 22761 Hamburg, Germany}
\author{Christian Reichl}
\affiliation{Laboratory for Solid State Physics, ETH Zürich, 8093 Zürich, Switzerland}
\author{Werner Wegscheider}
\affiliation{Laboratory for Solid State Physics, ETH Zürich, 8093 Zürich, Switzerland}
\author{Angel~Rubio}
 \affiliation{Max Planck Institute for the Structure and Dynamics of Matter, 22761 Hamburg, Germany}
\affiliation{Center for Computational Quantum Physics, Flatiron Institute, NY 10010 New York, USA}
\author{Jér\^{o}me Faist}
\affiliation{Institute of Quantum Electronics, ETH Zürich, 8093 Zürich, Switzerland}
\affiliation{Quantum Center, ETH Zürich, 8093 Zürich, Switzerland}

\maketitle

\section{Material and methods}\label{sec:matmet}
The two-dimensional electron system (2DES) is obtained in a GaAs/AlGaAs quantum well grown via molecular-beam epitaxy (MBE). The quantum well is \SI{30}{nm} wide and modulation doped, with a \SI{100}{nm} spacer. The electron density and mobility measured at \SI{1.3}{K} in the dark are $n_s = \SI{2.06e11}{cm^{-2}}$ and $\mu = \SI{1.59e7}{cm^2 V^{-1} s^{-1}}$, respectively. The sample is processed via standard photo-lithography techniques in clean-room environment. \\

The experiment is conducted with state-of-the-art methods for measurement of the quantum Hall effect in two-dimensional electron systems~\cite{baer2015transport}. We have employed a Bluefors dilution refrigerator capable of reaching temperatures down to \SI{10}{mK}. The voltage is measured with commercial MFLI Zurich Instruments digital lock-in amplifiers. Current is injected symmetrically in the Wheatstone bridge by applying an alternating-current (AC)-modulated voltage of \SI{1}{V} root-mean-square (rms) at the demodulation frequency $f_\mathrm{dem} = \SI{2.333}{Hz}$ to two \SI{100}{M \ohm} resistors in series with the Wheatstone bridge, such that a current of \SI{10}{nA} rms is circulating in the circuit. Differential AC low-noise pre-amplifiers are employed before the lock-in voltage input, so to amplify the signal by a factor of $1000$. The amplifier input voltage noise density amounts to \SI{2.6}{nV / \sqrt{Hz}} at the demodulation frequency $f_\mathrm{dem}$, while the input current noise is negligible in the balanced detection scheme. Moreover, the common mode amplification does not introduce any unwanted offset since in the balanced detection scheme with symmetric current injection the absolute voltages are close to zero. To demodulate the lock-in input signal we employ a fourth-order low-pass filter with time constant (TC) of \SI{1.0}{s}, which corresponds to a noise-equivalent-power (NEP) bandwidth of \SI{0.079}{Hz}, so that the input voltage noise is \SI{0.7}{nV}. The input noise of the lock-in signal input is negligible with respect to the noise of the amplified signal. Before the contacts to the sample, low-pass \SI{100}{kHz} filters are installed in order to minimize electric spikes or heating effects from the measurement setup.

\section{Wheatstone bridge circuital relation}
\begin{wrapfigure}{l}{0.5\textwidth}
    \includegraphics[width=\linewidth]{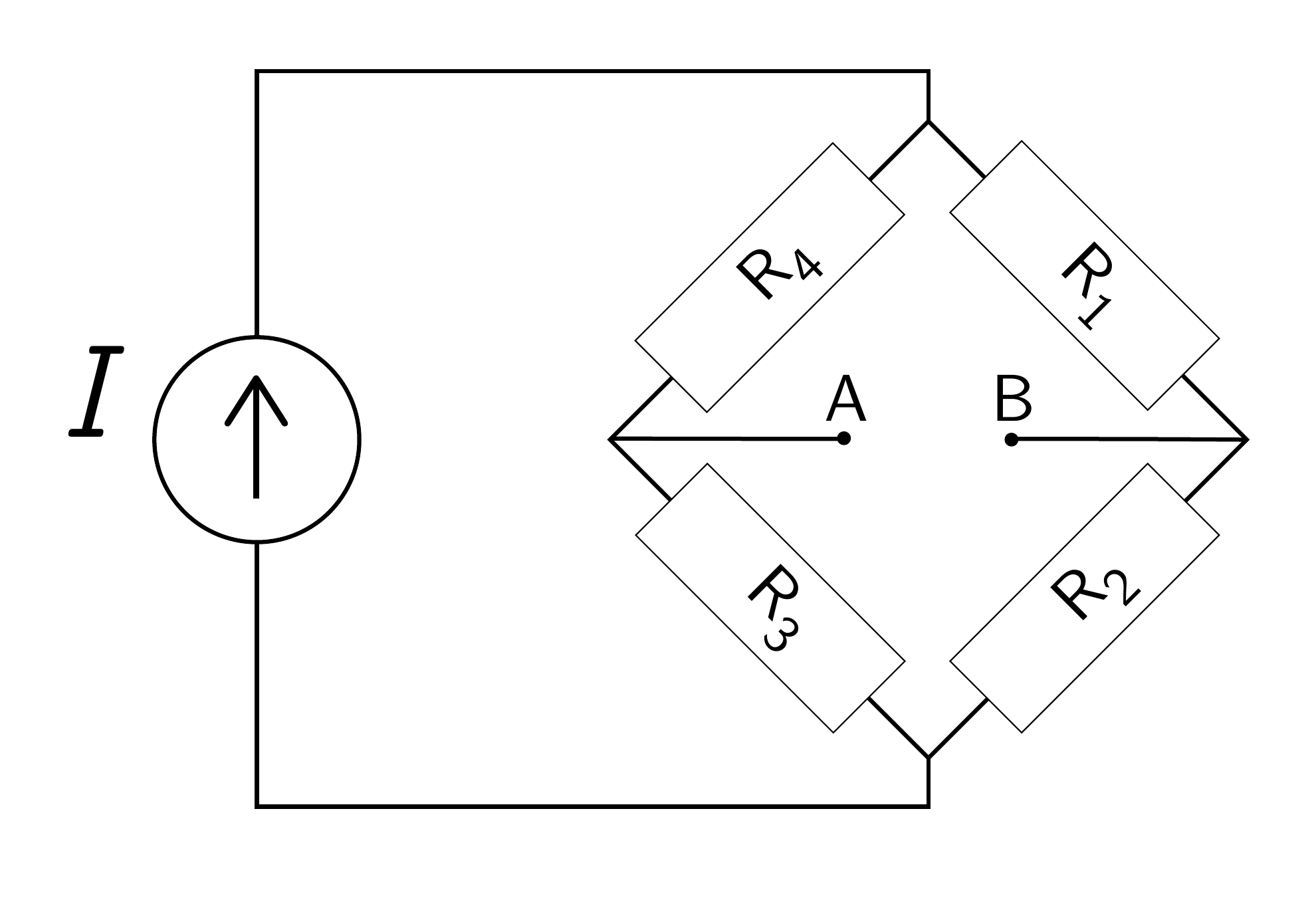}
    \caption{Circuital scheme of the Wheatstone bridge.}
    \label{fig:wheatstone_circuit}
\label{fig:wrapfig}
\end{wrapfigure}
In Figure~\ref{fig:wheatstone_circuit} we display the lumped-element circuit model of the Wheatstone bridge, from which one can obtain the relation given in Eq.~1 in the main text. By Kirchhoff's laws one obtains the unbalance voltage
\begin{equation}
    V_\mathrm{AB} = \frac{R_1 R_3 - R_4 R_2}{R_1 + R_2 + R_3 + R_4} I, 
\end{equation}
and assuming that all resistors but $R_4$ are equal, $R_1 = R_2 = R_3 = R_H$, while $R_4 = R_H + \Delta R_H$,
\begin{equation}
    \frac{\abs{V_\mathrm{AB}}}{I} = \frac{\Delta R_H}{4}.
\end{equation}
The measurement of the unbalance voltage is meaningful only at the quantum Hall plateaux, where the value of the resistance of the Hall bars not embedded in the cavity is the same and equal to $R_H = h / (e^2 \nu)$, with $\nu$ being the filling factor. By inverting the previous equation one then obtains
\begin{equation}
    \frac{\Delta R_H}{R_H} = 4 \frac{\abs{V_\mathrm{AB}}}{I} \frac{e^2}{h} \nu,
\end{equation}
which is Eq.~1 of the main text, where the unbalance voltage $\abs{V_\mathrm{AB}}$ is given by the average value $\langle V_\mathrm{ub} \rangle$ at the plateau of integer filling factor~$\nu$.

\section{Unbalance voltage traces of all samples}
\begin{figure}
    \centering
    \includegraphics[width=0.8\linewidth]{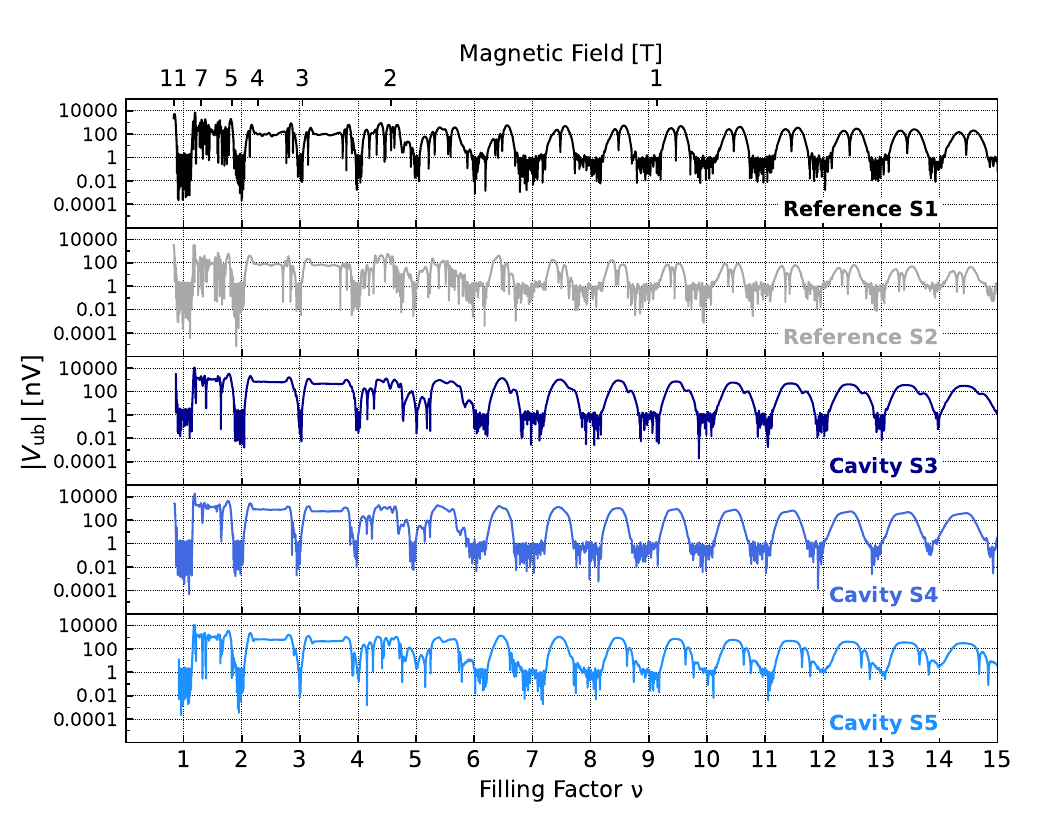}
    \caption{Absolute value of the unbalance voltage as a function of filling factor for the reference (S1-S2) and cavity (S3-S5) samples. Notice that the voltage axis is in logarithmic scale, to clarify how close is the value to zero at the integer filling factor plateaux.}
    \label{fig:allSamples_unbalance}
\end{figure}

\begin{figure}
    \centering
    \includegraphics[width=0.8\linewidth]{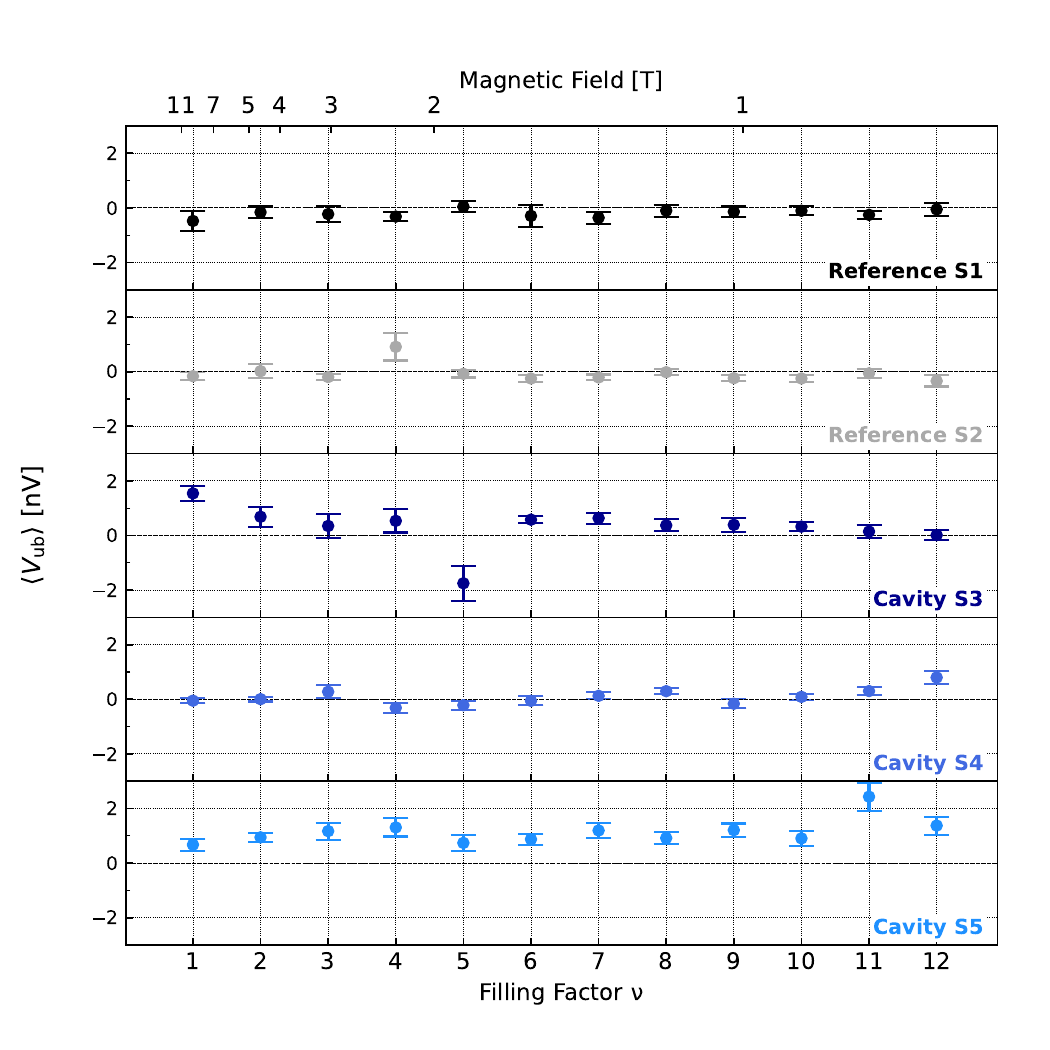}
    \caption{Average value of the unbalance voltage at the integer filling factor plateaux for the reference (S1-S2) and cavity (S3-S5) samples. The error-bars are given by the standard deviation of the values measured in the plateaux regions, and they have a similar length, given by the \SI{0.7}{nV} noise level.}
    \label{fig:allSamples_averages}
\end{figure}

In Figure~\ref{fig:allSamples_unbalance} we report the measurement of the in-phase quadrature of the unbalance voltage $V_\mathrm{ub}$ as a function of filling factor (that is of magnetic field, to be read on the top axis) of the reference (S1-S2) and cavity (S3-S5) samples. In order to highlight the values close to zero in the plateaux regions (at integer filling factors) we plot the absolute value of $V_\mathrm{ub}$ in logarithmic scale. As explained in the main text, the averaging of the unbalance voltage is performed at the integer filling factor plateaux, which are identified by taking the range of magnetic field in which $\log_{10}(\abs{V_\mathrm{ub}} / \SI{1}{V}) < -8.5$ (\SI{-170}{dB}), i.e.\ $\abs{V_\mathrm{ub}} < \SI{3.2}{nV}$. This choice is made to ensure that the plateau region is flat for all the samples, and allows to clearly distinguish its boundaries. Notice that the extension of the plateaux regions decreases with increasing filling factor, so that a plateau cannot be safely identified for filling factors greater than $12$. Moreover, we notice some variation in the extension of the plateaux also between different samples, which will reflect in the higher standard deviation of the average, as shown below. As already discussed in the main text, the differences in the values away from the plateaux regions arise due to the electronic density inhomogeneities of the 2DES in which the Hall bars are etched. 
In Figure~\ref{fig:allSamples_averages} we report the averages of the unbalance voltage for the integer filling factor plateaux. The error-bars represent the standard deviation of the mean keeping into account also systematic uncertainties, as discussed in Section~\ref{sec:uncertainties}. We notice that for both reference and cavity samples the values oscillate around zero, without showing any behaviour as a function of filling factor. Reference S1 and cavity S4 samples display the narrowest region of oscillation around zero, with a maximum distance of two standard deviations (i.e.\ they are compatible with zero within a 95\% confidence interval), except for filling factor 12 for S4, though cavity sample S3 is zero within one standard deviation at this filling factor. Since no deviation is observed for at least one cavity sample, this already provides an upper bound to the renormalized value of the von Klitzing constant. Reference S2 and cavity S3 and S5 samples show a larger range of oscillation around zero, and S5 in particular is consistently above zero, even if the distance is below three standard deviations (i.e.\ it is zero within a 99.7\% confidence interval). We ascribe this behaviour to the worse performance of the pre-amplifier channel employed to measure S5, which is introducing a larger phase error, thus the out-of-phase component partially leaks into the in-phase component (see Sec.~\ref{sec:uncertainties}), giving a finite positive offset. To a lower extent, also S3 displays a similar issue.

\section{Nature of the uncertainties}\label{sec:uncertainties}
There are two main sources of uncertainty which limit the precision and the accuracy of the experiment: one is statistical, and it is given by the fluctuations of the voltage in the plateaux regions due to the voltage noise of the AC pre-amplifier, the other one is systematic, and it arises due to inaccuracies of the measurement setup, or to sample-specific characteristics, such as the different contact resistances, or capacitative effects arising in the 2DES. Assuming that the uncertainty sources are independent, we can obtain, for each plateau, the total uncertainty on the unbalance voltage (expressed with the standard deviation, which is represented via error-bars in Fig.~\ref{fig:allSamples_averages}) as $\sigma_V = \sqrt{\sigma_\mathrm{noise}^2 + \sigma_\mathrm{offset}^2 + \sigma_\mathrm{sample}^2}$, and we proceed to discuss the meaning of the terms in the following.

The term $\sigma_\mathrm{noise}$ is the standard deviation of the average value of the unbalance voltage in the plateau, and it is statistical in nature. If successive data points were uncorrelated it would simply be $\sigma_\mathrm{noise} = \sigma_\mathrm{noise}^\mathrm{uncorr} = \sqrt{\var{V_\mathrm{ub}} / N}$, where $N$ is the number of data points acquired in the plateau, and $\var{V_\mathrm{ub}}$ is their sample variance, which is equal to the \SI{0.7}{nV} pre-amplifier voltage noise (see Section~\ref{sec:matmet}). However, the data display some correlation due to the fact that the magnetic field is increased at a rate between \SI{10}{} and \SI{50}{mT / min} (increasing with increasing field), and the fourth-order low pass filter employed for lock-in demodulation has a settling time equal to about 10 times TC, that is \SI{10}{s}. Hence we perform a blocking statistical analysis to properly take into account the correlations, and we obtain an uncertainty which is about an order of magnitude greater than $\sigma_\mathrm{noise}^\mathrm{uncorr}$ and, for filling factors greater or equal to $3$, gives the dominant contribution to the total uncertainty.

The term $\sigma_\mathrm{offset}$ represents the offset which may be introduced by inaccuracies in the measurement setup. It is systematic in nature, and to assess it we measure the unbalance voltage of a short positioned in place of the sample. We obtain an estimate of $\sigma_\mathrm{offset} = \SI{0.07}{nV}$. The pre-amplifier channel could however also introduce a slight phase error, which combined with a non-zero out-of-phase component of the signal (see below) could give rise to a finite offset in the in-phase component. 

\begin{figure}[hbt]
    \centering
    \includegraphics[width=0.8\linewidth]{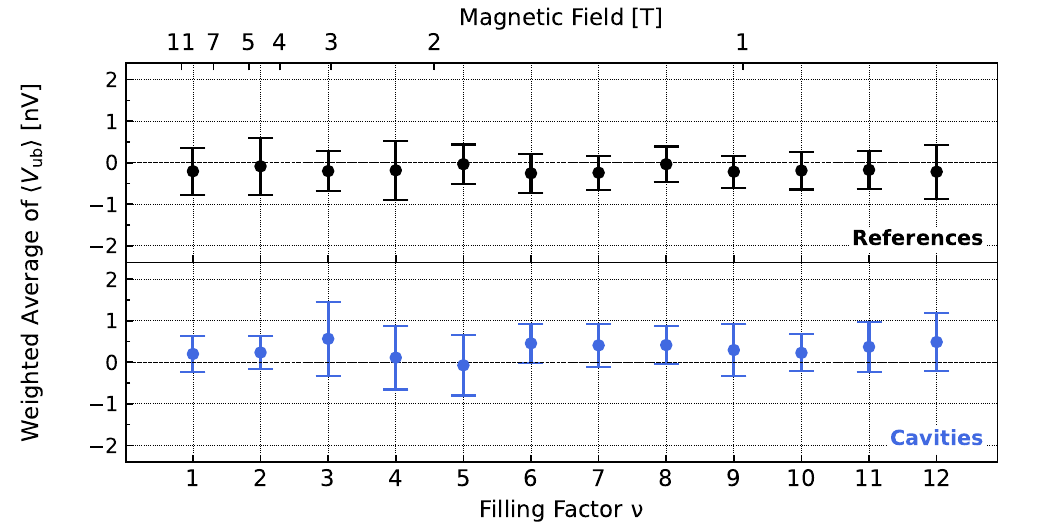}
    \caption{Weighted averages of the unbalance voltage for reference S1-S2 samples (top panel) and cavity S3-S5 samples (bottom panel), as a function of filling factor. The error-bars are given here by three times the (weighted) standard deviation, i.e.\ they refer to a 99.7\% confidence interval.}
    \label{fig:weighted_avg_Vub}
\end{figure}

Finally, the term $\sigma_\mathrm{sample}$ is the systematic uncertainty arising from sample-specific characteristics. As already discussed in the main text, the multiple connection technique~\cite{multipleConnections} allows to reduce the relative inaccuracy introduced by the different contact resistances down to $(R_C / R_H)^4$, where $R_H = h / (e^2 \nu)$ is the quantized Hall resistance of the plateau, and $R_C$ is the contact resistance, which varies between \SI{200}{} and \SI{300}{\ohm} (measured via the three-terminal measurement scheme). Due to capacitative effects of the 2DES the out-of-phase quadrature of the signal $V_Y$ is different from zero in the plateau regions, with values between \SI{2}{} and \SI{10}{nV} (increasing with decreasing filling factor). Being it out-of-phase it should not bear any impact on the in-phase quadrature of the unbalance voltage $V_X$ (displayed in Fig.~\ref{fig:allSamples_unbalance}), however being the in-phase component much lower than the out-of-phase one, the latter can partially contribute to a non-zero value of the former due to the slight phase shift introduced by the pre-amplifier channels. We assess this uncertainty from the reference samples, where we estimate a deviation of the out-of-phase component from the $90$ degrees phase up to $10$ degrees, so that the in-phase component $V_X$ can acquire an offset of $\sqrt{V_X^2 + V_Y^2}\sin{(10^\circ)}$, which we adopt as $\sigma_\mathrm{sample}$. This uncertainty contribution dominates at the plateaux at filling factors 1 and 2, and we interpret it as an higher reactance of the contacts to the 2DES at higher magnetic fields. The choice of measuring more than one reference/cavity sample (on the same chip) is thus suitable to assess the sample-specific uncertainties, and to statistically treat them. We display in Figure~\ref{fig:weighted_avg_Vub} the weighted mean of both reference and cavity samples, along with their standard deviation (the error-bars are three standard deviation-long, so to display the 99.7\% confidence interval). The weighted mean $\overline{\langle V_\mathrm{ub} \rangle}$ and the standard deviation of the weighted mean $\sigma_{\bar{V}}$ are defined as
\begingroup
\large
\begin{equation}
    \overline{\langle V_\mathrm{ub} \rangle} = \frac{\sum_i \frac{\langle V_\mathrm{ub} \rangle_i}{ (\sigma_V)_{i}^2}}{\sum_i \frac{1}{(\sigma_V)_{i}^2}}, \quad\quad \sigma_{\bar{V}} = \sqrt{\frac{1}{\sum_i \frac{1}{(\sigma_V)_{i}^{2}}}},
\end{equation}
\endgroup
where the sums are extended over all the reference/cavity samples.
We can infer that no deviation is observed within a range of \SI{0.7}{nV} for filling factor 1, which translates to a precision of 1 part in $10^5$ for the Hall resistance deviation, as discussed in the main text.

\section{Theoretical background}
\subsection{Homogeneous 2D Quantum Hall System Coupled to the Cavity}

We consider a homogeneous, two-dimensional electron gas coupled to a strong magnetic field and a single-mode homogeneous cavity field. The system is described by the Pauli-Fierz Hamiltonian~\cite{rokaj2017, cohen1989photons}
\begin{equation}
\hat{H} = \sum^{N}_{i=1}\frac{\big(\bm{\pi}_i+e\hat{\bi{A}}\big)^2}{2m}
 + \hbar\omega_{\rm{cav}}\left(\hat{a}^{\dagger}\hat{a} + \frac{1}{2}\right) + \sum_{i<j}W(\bi r_i-\bi r_j),\label{Pauli-Fierz}
\end{equation}
where $\bm{\pi}_i=\textrm{i}\hbar\bm\nabla_i+e\bi{A}_{\textrm{ext}}(\bi{r}_i)$ are the dynamical momenta of the electrons and $\mathbf{A}_{\textrm{ext}}(\mathbf{r})=-\bi{e}_xBy$ describes the homogeneous magnetic field $\bi{B}=\bm\nabla \times \bi{A}_{\textrm{ext}}(\bi{r})=B\bi{e}_z$. The cavity field $\hat{\bi{A}} = \sqrt{\frac{\hbar}{2\epsilon_0\mathcal{V}\omega_{\rm{cav}}}}\bi{e}_x\left(\hat{a} + \hat{a}^{\dagger}\right)$ is characterized by the in-plane polarization vector $\bi{e}_x$ and the cavity frequency $\omega_{\rm{cav}}$. The $\mathcal{V}$ and $\epsilon_0$ are the effective mode volume and the dielectric constant, respectively, and the ladder operators $\hat{a}$ and $\hat{a}^{\dagger}$ represent the bare photon fields. With Galilean invariance in a purely homogeneous system, the center of mass (CM) is decoupled from the relative motion of the electrons, regardless of the interaction strength~\cite{KohnMode}. To demonstrate this we use the CM and relative distance coordinates
\begin{equation}
    \mathbf{R}=\frac{1}{\sqrt{N}}\sum^N_{i=1}\mathbf{r}_i \;\; \textrm{and}\;\; \widetilde{\mathbf{r}}_{j}=\frac{\mathbf{r}_1-\mathbf{r}_j}{\sqrt{N}}\;\;\textrm{with}\;\;j>1.
\end{equation}
The Hamiltonian in the new frame is the sum of two parts: (i) the CM part $\hat{H}_{\textrm{cm}}$ which is coupled to the quantized field $\hat{\mathbf{A}}$ and (ii) the relative distances $\hat{H}_{\textrm{rel}}$ which does not couple to the cavity field, $\hat{H}=\hat{H}_{\textrm{cm}}+\hat{H}_{\textrm{rel}}$
where each part looks as
\begin{eqnarray}
 \hat{H}_{\textrm{cm}}&=&\frac{1}{2m}\left(\textrm{i}\hbar\nabla_{\mathbf{R}}+e\mathbf{A}_{\textrm{ext}}(\mathbf{R})+e\sqrt{N}\hat{\mathbf{A}}\right)^2+\hbar\omega_{\rm{cav}}\left(\hat{a}^{\dagger}\hat{a}+\frac{1}{2}\right)\\
  \hat{H}_{\textrm{rel}}&=&\frac{1}{2m}\sum^N_{j=2}\left(\frac{\textrm{i}\hbar}{\sqrt{N}}\widetilde{\nabla}_{j}+e\sqrt{N}\mathbf{A}_{\textrm{ext}}(\widetilde{\mathbf{r}}_j)\right)^2 -\frac{\hbar^2}{2mN}\sum^N_{j,l=2}\widetilde{\nabla}_{j}\cdot\widetilde{\nabla}_{l}-\frac{e^2}{2m}\left(\sum^N_{j=2}\mathbf{A}_{\textrm{ext}}(\widetilde{\mathbf{r}}_j)\right)^2+\sum_{i<j}W(\bi r_i-\bi r_j).
\end{eqnarray}
From the result above it becomes evident that in the homogeneous limit the quantum cavity field only couples to the CM of the electron gas while the correlations decouple from it. This implies that all the cavity-matter phenomena happen at the CM.

\subsection{Collective Landau Polaritons}

 The Hamiltonian $\hat{H}_{\rm cm}$ has the form of two coupled harmonic oscillators, one for the Landau level transition and one for the photons. In many cases such a Hamiltonian is known as the Hopfield Hamiltonian which can be solved by the Hopfield transformation~\cite{Hopfield}. After the Hopfield transformation we find
\begin{equation}
    \hat{H}_{\rm cm} = \hbar\Omega_+\left(\hat{b}^\dag_+\hat{b}_++\frac12\right) + \hbar \Omega_-\left(\hat{b}^\dag_-\hat{b}_-+\frac12\right)
\end{equation}
where $\{\hat{b}^{\dagger}_{\pm},\hat{b}_{\pm}\}$ are the creation and annihilation operators of the Landau polariton quasiparticles. The $\Omega_{\pm}$ are the upper and lower Landau polariton modes respectively,
\begin{eqnarray}\label{Polariton modes}
 \Omega^2_{\pm}=\frac{\omega^2_{\rm{cav}}+\omega^2_d+\omega^2_c}{2}\pm\sqrt{\omega^2_d\omega^2_c+\left(\frac{\omega^2_{\rm{cav}}+\omega^2_d-\omega^2_c}{2}\right)^2}
\end{eqnarray} 
where $\omega_d=\sqrt{e^2N/m_e\epsilon_0 \mathcal{V}}$ is the diamagnetic frequency originating from the $\hat{\mathbf{A}}^2$ which depends on the number of electrons $N$ and the effective mode volume $\mathcal{V}$. To define the polariton operators we represent the photon annihilation operator in terms of a displacement coordinate $q$ and its conjugate momentum as $\hat{a} = (q+\partial_q)/\sqrt2$, with $\hat{a}^{\dagger}$ obtained via conjugation. In this representation the polariton operators $\{\hat{b}_{\pm},\hat{b}^{\dagger}_{\pm}\}$ can be written in terms of mixed, polaritonic coordinates as $S_{\pm}=\sqrt{\hbar/2\Omega_{\pm}}\left(\hat{b}_{\pm}+\hat{b}^{\dagger}_{\pm}\right)$ with
\begin{eqnarray}
    S_+ = \frac{\sqrt{m_e} \bar Y+q\Lambda\sqrt{\hbar/\omega_{\rm{cav}}}}{\sqrt{1+\Lambda^2}}\; \textrm{and}\; S_- = \frac{-q\sqrt{\hbar/\omega_{\rm{cav}}} +\sqrt{m_e} \Lambda \bar Y}{\sqrt{1+\Lambda^2}}\nonumber
\end{eqnarray}
where $\bar Y=Y+\frac{\hbar K_x}{eB}$ is the guiding center and $K_x$ is the electronic wave number in the $x$-direction. Also we introduced the parameter $\Lambda=\alpha-\sqrt{1+\alpha^2}$ with $ \alpha=\left(\omega^2_c-\omega^2_{\rm{cav}}-\omega^2_d\right)/2\omega_d\omega_c$ which quantifies the mixing between electronic and photonic degrees of freedom.

\subsection{Finite Temperature Kubo Transport}

Here we present the Kubo linear-response formalism for the finite temperature transport of the light-matter system. As we already showed the Hamiltonian of our system can be written as a sum of a CM and relative part $\hat{H}=\hat{H}_{\rm cm}+\hat{H}_{\rm rel}$. To proceed we assume that the eigenstates of $\hat{H}_{\rm cm}$ are $|\Phi_n\rangle$ and the eigenstates of $\hat{H}_{\rm rel}$ are $|F_{I}\rangle$ such that it holds
\begin{eqnarray}
\hat{H}_{\rm cm}|\Phi_{n}\rangle=E_n|\Phi_n\rangle \;\; \textrm{and}\;\; \hat{H}_{\rm rel}|F_{I}\rangle=E_{I}|F_{I}\rangle
\end{eqnarray}
Then, the eigenstates of the full Hamiltonian $\hat{H}$ are $ |\Psi_{nI}\rangle=|\Phi_n\rangle\otimes |F_{I}\rangle$, and the full eigenspectrum is $E_{nI}=E_{n}+E_{I}$. The Kubo formula for the optical conductivity of the system is~\cite{kubo, ALLEN2006165}
\begin{eqnarray}\label{Conductivities}
 \sigma_{ab}(w)=\frac{\textrm{i}}{w+\textrm{i}\delta}\left(\frac{e^2n_e}{m_e}\delta_{ab}+\frac{\chi_{ab}(w)}{A}\right)\;\;\; \delta \rightarrow 0^+
\end{eqnarray}
where $a,b=x,y,z$. The first term in the optical conductivity is the Drude term, while the second term is the current-current correlator in the frequency domain, which is defined as the Fourier transform of current-current correlator in the time domain
\begin{eqnarray}
     \chi_{ab}(t)=\frac{-\textrm{i}\Theta(t)}{\hbar }\langle[\hat{J}_{a}(t),\hat{J}_{b}]\rangle.
 \end{eqnarray}
The current operators are considered in the interaction picture $\hat{\mathbf{J}}(t)=e^{\textrm{i}Ht/\hbar}\hat{\mathbf{J}}e^{-\textrm{i}\hat{H}t/\hbar}$~\cite{kubo}. In the canonical ensemble the expectation value of an operator $\hat{\mathcal{O}}$ is defined as~\cite{ALLEN2006165}
\begin{eqnarray}
    \langle\hat{\mathcal{O}}\rangle=Tr\{\hat{\rho} \hat{\mathcal{O}}\}=\frac{1}{\mathcal{Z}}\sum_{n,I} \langle \Psi_{nI}|e^{-\beta \hat{H}}\hat{\mathcal{O}}|\Psi_{nI}\rangle
\end{eqnarray}
where the partition function is $\mathcal{Z}=\sum_{n,I}e^{-\beta E_n}e^{-\beta E_{I}}$. We will use these formulas now for the computation of the current correlation functions. The current response can be splitted into two parts 
\begin{eqnarray}\label{JJResponse}
     \chi_{ab}(t)=\frac{-\textrm{i}\Theta(t)}{\hbar }\left(\langle\hat{J}_a(t)\hat{J}_b\rangle -\langle \hat{J}_b\hat{J}_a(t)\rangle \right).
 \end{eqnarray}
Let us compute first the first term $\langle\hat{J}_a(t)\hat{J}_b\rangle$. We use the expression for the canonical ensemble and for the current operator in the interaction picture and we have
\begin{eqnarray}
    \langle\hat{J}_a(t)\hat{J}_b\rangle&=&\frac{1}{\mathcal{Z}}\sum_{n,I}e^{-\beta E_{nI}} \langle \Psi_{nI}|e^{\textrm{i}Ht/\hbar}\hat{J}_ae^{-\textrm{i}\hat{H}t/\hbar} \hat{J}_b|\Psi_{nI}\rangle=\frac{1}{\mathcal{Z}}\sum_{n,I}e^{-\beta E_{nI}} e^{\textrm{i}tE_{nI}/\hbar} \langle \Psi_{nI}|\hat{J}_ae^{-\textrm{i}\hat{H}t/\hbar} \hat{J}_b|\Psi_{nI}\rangle.
\end{eqnarray}
We introduce the identity $\mathbb{I}=\sum_{m,J}|\Psi_{mJ}\rangle\langle \Psi_{mJ}|$ in the above expression 
\begin{eqnarray}
  \langle\hat{J}_a(t)\hat{J}_b\rangle&=& \frac{1}{\mathcal{Z}}\sum_{n,m,J,I}e^{-\beta E_{nI}} e^{\textrm{i}tE_{nI}/\hbar} \langle \Psi_{nI}|\hat{J}_ae^{-\textrm{i}\hat{H}t/\hbar} |\Psi_{mJ}\rangle\langle \Psi_{mJ}|\hat{J}_b|\Psi_{nI}\rangle \nonumber \\
  &=& \frac{1}{\mathcal{Z}}\sum_{n,m,J,I}e^{-\beta E_{nI}} e^{\textrm{i}t(E_{nI}-E_{mJ})/\hbar}  \langle \Psi_{nI}|\hat{J}_a|\Psi_{mJ}\rangle\langle \Psi_{mJ}|\hat{J}_b|\Psi_{nI}\rangle 
\end{eqnarray}

Since we work in the CM frame in order to proceed we need examine how the current operator looks in the CM frame. The full gauge-invariant current operator in the original frame is~\cite{Landau}
\begin{eqnarray}\label{Current Operator}
    \hat{\bi{J}}=-\frac{\textrm{i}e\hbar}{m_{\textrm{e}}}\sum^N_{j=1}\nabla_j-\frac{e^2N}{m_{\textrm{e}}}\hat{\bi{A}}-\frac{e^2}{m_{\textrm{e}}}\sum^N_{i=1}\bi{A}_{\textrm{ext}}(\mathbf{r}_i).
\end{eqnarray}
Then, the current operator in the CM frame takes the form
\begin{equation}\label{Current COM}
    \hat{\mathbf{J}}=\sqrt{N}\left[-\frac{\textrm{i}e\hbar}{m_e}\nabla_{\mathbf{R}}-\frac{e^2}{m_e}\sqrt{N}\hat{\mathbf{A}}-\frac{e^2}{m_e}\mathbf{A}_{\textrm{ext}}(\mathbf{R})\right]\equiv \hat{\mathbf{J}}_{\rm cm}.
\end{equation} 
The above result shows that the total current in the system is equal to current of the CM and depends only on CM related operators. This property has the following important implication
\begin{equation}
    \langle \Psi_{nI}|\hat{\mathbf{J}}|\Psi_{mJ}\rangle=\delta_{IJ} \langle \Phi_n|\hat{\mathbf{J}}|\Phi_m\rangle
\end{equation}
using the above the expression for the current correlator simplifies 
\begin{eqnarray}
  \langle\hat{J}_a(t)\hat{J}_b\rangle=\frac{1}{\mathcal{Z}}\sum_{n,m,I}e^{-\beta E_{nI}} e^{\textrm{i}t(E_{n}-E_{m})/\hbar}  \langle \Phi_n|\hat{J}_{a}|\Phi_m\rangle\langle \Phi_{m}|\hat{J}_{b}|\Phi_{n}\rangle\nonumber\\ 
\end{eqnarray}
To obtain the above we used that $E_{nI}-E_{mI}=E_{n}-E_{m}$. To complete the computation we need to multiply $\langle\hat{J}_a(t)\hat{J}_b\rangle$ with $\frac{-\textrm{i}\Theta(t)}{\hbar }$ and Fourier transform into the frequency space
\begin{eqnarray}
    \frac{-\textrm{i}\Theta(t)}{\hbar } \langle\hat{J}_a(t)\hat{J}_b\rangle & \longrightarrow & \frac{1}{\mathcal{Z}}\sum_{n,m,I}e^{-\beta E_{nI}}   \frac{\langle \Phi_n|\hat{J}_{a}|\Phi_m\rangle\langle \Phi_{m}|\hat{J}_{b}|\Phi_{n}\rangle}{w+(E_n-E_m)/\hbar +\textrm{i}\delta}=\frac{\sum_{I}e^{-\beta E_{I}}}{\sum_{I}e^{-\beta E_{I}}\sum_{k}e^{-\beta E_k}} \sum_{n,m,I}e^{-\beta E_{n}}   \frac{\langle \Phi_n|\hat{J}_{a}|\Phi_m\rangle\langle \Phi_{m}|\hat{J}_{b}|\Phi_{n}\rangle}{w+(E_n-E_m)/\hbar +\textrm{i}\delta}\nonumber \\
    &&=\frac{1}{\sum_{k}e^{-\beta E_k}} \sum_{n,m}e^{-\beta E_{n}}   \frac{\langle \Phi_n|\hat{J}_{a}|\Phi_m\rangle\langle \Phi_{m}|\hat{J}_{b}|\Phi_{n}\rangle}{w+(E_n-E_m)/\hbar +\textrm{i}\delta} \;\; \textrm{with}\;\; \delta \rightarrow 0^+.
\end{eqnarray}
Following exactly the same procedure for the second term in Eq.(\ref{JJResponse}) $\frac{\textrm{i}\Theta(t)}{\hbar } \langle \hat{J}_b\hat{J}_a(t)\rangle$ we find the the expression for the current-current response function
\begin{equation}\label{Tresponse}
    \chi_{ab}(w)=\frac{1}{\sum_{l}e^{-\beta E_l}} \sum_{n,m}\left(e^{-\beta E_{n}}-e^{-\beta E_m}\right)   \frac{\langle \Phi_n|\hat{J}_{a}|\Phi_m\rangle\langle \Phi_{m}|\hat{J}_{b}|\Phi_{n}\rangle}{w+(E_n-E_m)/\hbar +\textrm{i}\delta} \;\; \textrm{with}\;\; \delta \rightarrow 0^+.
\end{equation}
From the above expression we see that current response function solely depends on the CM eigenstates and the CM eigenenergies. This is a consequence of homogeneity which implies the separability of the full Hamiltonian into CM and relative parts. Finally, for completeness we provide the expressions for the components of the current operator in terms of the polaritonic annihilation and creation operators
\begin{eqnarray}
    \hat{J}_x&=&\frac{e^2\sqrt{N}B}{m^{3/2}_e}\sqrt{\frac{\hbar}{2(1+\Lambda^2)}}\left[\frac{\sqrt{m_e}}{eB}\left(-\textrm{i}\hbar\nabla_X-\hbar K_x\right)+\frac{\Lambda+\eta}{\sqrt{\Omega_-}}\left(\hat{b}^{\dagger}_-+\hat{b}_-\right)+\frac{1-\eta\Lambda}{\sqrt{\Omega_+}}\left(\hat{b}^{\dagger}_++\hat{b}_+\right)\right] \nonumber\\
\hat{J}_y&=&-\textrm{i}e\sqrt{\frac{ \hbar N}{2m_e(1+\Lambda^2)}}\left[\sqrt{\Omega_+}\left(\hat{b}_+-\hat{b}^{\dagger}_+\right)+\Lambda\sqrt{\Omega_-}\left(\hat{b}_--\hat{b}^{\dagger}_-\right)\right].
\end{eqnarray}
With the above expressions one can use straightforwardly apply the Kubo formalism and compute the temperature dependent current response functions in Eq.~(\ref{Tresponse}) and the corresponding conductivities defined in Eq.~(\ref{Conductivities}). For more details on the finite temperature Kubo formalism for the Landau polariton states one can see Ref.~\cite{rokaj2023topological}.

\section{No Modification of Quantum Hall Transport at $T=0$}

Having derived the general formula for the current correlator $\chi_{ab}(w)$ at finite temperature, we will focus now at the transport properties at zero temperature, $T=0$, where the topological protection and the quantization of the quantum Hall conductance are expected from the Thouless argument, as long as the system is gapped~\cite{PhysRevLett.49.405}. At $T=0$ the ground state of the polariton system is for $n_+=n_-=0$ and only the thermal prefactors corresponding to the ground state $e^{-\beta E_{00}}$ contribute to transport. 
\begin{equation}
    \chi_{ab}(w)=\sum_{m_+,m_-} \frac{\langle 00|\hat{J}_{a}|m_+m_-\rangle\langle m_+m_-|\hat{J}_{b}|00\rangle}{w+(E_{00}-E_{m_+m_-})/\hbar +\textrm{i}\delta} - (00 \leftrightarrow m_+m_-)
\end{equation}
Furthermore, the current operators are linear in the polaritonic annihilation and creation operators and thus allow only for single-polariton transitions to occur, which implies that in the denominator of the response function only the single polariton energies $\Omega_{\pm}$ show up. Finally, using the formulas for the matrix representation of the components of the current operator we find the following analytically exact expressions for the transverse $\chi_{xy}$ response function
\begin{eqnarray}
 \chi_{xy}(w)=\frac{Ne^3B}{(1+\Lambda^2)m^2_e}\Big[ \Lambda(\Lambda+\eta)\frac{\textrm{i}}{2}\left(\frac{1}{w+\Omega_-+\textrm{i}\delta}+\frac{1}{w-\Omega_-+\textrm{i}\delta}\right)
 +(1-\eta\Lambda)\frac{\textrm{i}}{2}\left(\frac{1}{w+\Omega_++\textrm{i}\delta}+\frac{1}{w-\Omega_++\textrm{i}\delta}\right)\Big].
\end{eqnarray}
With the above result and using the Kubo formula in Eq.~(\ref{Conductivities}) we find the Hall conductance by taking the limit $w \rightarrow 0$
\begin{eqnarray}
      \sigma_{xy}=\frac{e^2\nu}{h(1+\Lambda^2)}\left[\frac{\Lambda(\Lambda+\eta)}{\Omega^2_-/\omega^2_c+\delta^2/\omega^2_c}+\frac{1-\eta\Lambda}{\Omega^2_+/\omega^2_c+\delta^2/\omega^2_c}\right] \;\; \textrm{where}\;\; \eta=\omega_d/\omega_c.
\end{eqnarray}
Where we introduced the Landau level filling factor $\nu=n_sh/eB$, as defined in the main text. Taking the value of the broadening parameter to zero $\delta \rightarrow 0$ we find  that the Hall conductance to be quantized
\begin{equation}
    \sigma_{xy}=\frac{e^2\nu}{h},
\end{equation}
consistent with the Thouless flux insertion argument~\cite{PhysRevLett.49.405}. In the last step we used two properties of the mixing parameter $
 1- \eta \Lambda = \Omega^2_+/\omega^2_c\;\;\textrm{and}\;\; \Lambda(\Omega^2_-/\omega^2_c-1)=\eta
$ which can be exactly deduced from the definition of $\Lambda$. 

\subsection{Singular Point}
We would also like to comment on the limit where the cavity frequency $\omega_{\rm{cav}}$ is much smaller than the other energy scales in the light-matter system, namely the diamagnetic frequency $\omega_d$, and the cyclotron frequency $\omega_c$, i.e., $\omega_{\rm{cav}} \ll \omega_d, \omega_c$. In this limit where the cavity frequency becomes negligible it can be easily seen that the lower polariton mode goes to zero, $\Omega_- \rightarrow 0$, as it was found in~\cite{rokaj2019}. Having the lower polariton gap closing circumvents the Thouless flux insertion argument~\cite{PhysRevLett.49.405} which assumes the system to be gapped. The gap closing leads to the modification of the Hall conductance as it was found in Ref.~\cite{rokaj2022polaritonic}.

\bibliography{bibliography}